\documentclass[authoryear,12pt,3p]{jowarticle}
\usepackage{graphicx}
\usepackage{amsmath}
\usepackage{amssymb}
\usepackage{amsfonts}
\usepackage{amsthm}
\usepackage{natbib}
\usepackage{setspace}
\usepackage[all]{xy}
\usepackage{enumitem}
\usepackage{titlesec}
\usepackage{mathrsfs}
\makeatletter
\renewcommand\section{\@startsection{section}{1}{\z@}{-3.25ex plus -1ex minus -.2ex}{1.5ex plus .2ex}{\normalsize\bf}}
\renewcommand\subsection{\@startsection{subsection}{2}{\z@}{-3.25ex plus -1ex minus -.2ex}{1.5ex plus .2ex}{\normalsize\bf}}
\renewcommand\subsubsection{\@startsection{subsubsection}{3}{\z@}{-3.25ex plus -1ex minus -.2ex}{1.5ex plus .2ex}{\normalsize\bf}}
\makeatother

\newtheorem{thm}{Theorem}

\newtheorem{lem}[thm]{Lemma}
\newtheorem{prop}[thm]{Proposition}

\begin{document}
\begin{frontmatter}
\title{(Information) Paradox Regained? \\ A Brief Comment on Maudlin on Black Hole Information Loss}
\author{JB Manchak}\ead{jmanchak@uci.edu}
\author{James Owen Weatherall}\ead{weatherj@uci.edu}
\address{Department of Logic and Philosophy of Science\\ University of California, Irvine}

\begin{abstract}
We discuss some recent work by Tim Maudlin concerning Black Hole Information Loss.  We argue, contra Maudlin, that there is a paradox, in the straightforward sense that there are propositions that appear true, but which are incompatible with one another.  We discuss the significance of the paradox and Maudlin's response to it.
\end{abstract}

\end{frontmatter}

The \emph{black hole information loss paradox} \citep{HawkingIL} is a widely discussed (putative) puzzle that arises when one attempts to make sense of physicists' expectation that global quantum dynamics will be unitary in light of the postulated phenomenon of black hole evaporation \citep{HawkingR1,HawkingR2,UnruhBHE}, which is apparently a consequence of Hawking radiation.\footnote{See \citet{Unruh+Wald} for a recent (opinionated) review of the issues and the literature over the last four decades.}  In a provocative recent manuscript, Tim \citet{Maudlin} forcefully argues that ``There is no `information loss' paradox.''  The ``solution'' to the ``paradox'', he claims, requires no new physics and indeed, was already available, though not appreciated, in 1975.  The appearance of paradox arises only because of persistent errors, both mathematical and conceptual in character, by prominent members of the physics community.

There is much to admire in Maudlin's treatment of the subject.  We agree, for instance, with his discussion of the (non-)role of ``information'' in the putative paradox.  What is really at issue is predictability, or even better, retrodictability, and not information (at least in the sense of information theory): as we will describe in more detail below, the puzzle concerns whether any specification of data on a particular surface is sufficient to retrodict the physical process by which that data came about.  We also think that Maudlin draws helpful attention to the core issues, which sometimes get obscured in physicists' discussions of solutions to the paradox. For instance, as Maudlin correctly observes, one can only expect unitary evolution between states defined on a special class of surfaces, known as \emph{Cauchy surfaces}; any conceptually adequate discussion of black hole information loss will need to consider the conditions under which suitable Cauchy surfaces are available in the first place.

Still, we do not agree with his central claim.  As we will presently argue, we take it that there \emph{is} a paradox, in the precise sense that there are well-motivated and widely accepted assertions that can be shown to be mutually inconsistent.\footnote{Is this really a paradox?  \citet{lycan2010exactly} offers an extension of an account of paradoxes due to \citet{Quine} according to which cases of apparently-true-but-mutually-inconsistent premises  certainly do count as paradoxes (though it is not clear that they fit into Quine's original classification). Other authors do not call such cases ``paradoxes''; for instance, \citet{mcgrath2007four} calls them ``problem sets''.  We are not particularly committed to the term ``paradox'', but we do think the expression ``black hole information loss problem set'' invokes too many memories of graduate school and so we stick with the standard usage.}  We take it, then, that any ``solution'' to the paradox will require one to reject one of these assertions.  In other words, although the literature may well be rife with mathematical and conceptual errors, merely identifying these errors is not sufficient to render the assertions in question consistent.  One still must give something up---as, we will argue, Maudlin ultimately does.  Indeed, it seems to us that rather than showing that there is no paradox, Maudlin endorses a radical solution to the paradox on which one rejects the assumption that space and time can be represented by a smooth Lorentzian manifold (with or without boundary).  One is certainly entitled to resolve the paradox in this way, but it is important to recognize that there are real costs involved.  In particular, if the motivating concern behind black hole information loss was a failure of retrodictability, then the solution Maudlin offers is pyrrhic.\footnote{We do not claim that what we describe here is the only ``information loss paradox.''  \citet{Wallace}, for instance, emphasizes a different paradox related to black hole evaporation, and points out that there are several distinct puzzles in the physics literature that go under this name.  Still, the paradox that we identify here seems to us to be the core of what is at issue in \citet{Maudlin}---and also, for instance, in \citet{Unruh+Wald}.}

We admit that there is a certain sense in which (almost) everything we say appears already in Maudlin's paper.  Indeed, in many ways we follow his treatment.  Our disagreement, perhaps, is a matter of emphasis.  But there is at least one place where we think Maudlin's discussion is misleading, and in any case, it seems to us that the differences of emphasis are themselves worth emphasizing.  And so we submit a different perspective.

As a first pass, we will introduce and discuss two assertions that we take to be well-motivated and widely accepted, but which can be shown to be inconsistent.  (Showing the inconsistency is not novel: it was established by \citet{Kodama}, though our treatment will follow \citet{Wald1984}; see also \citet{Earman}.  Still, perhaps the result has not been appreciated.)  This will be our statement of the paradox; standard replies to the paradox amount to rejecting one or the other of these assertions.  After presenting this discussion, we will circle back to Maudlin's argument, and present what we take to be his solution to the paradox.  As we will argue, Maudlin's response does not involve rejecting either of the premises we will state.  Instead, as we said above, he rejects a suppressed background assumption, concerning the continuity of a Lorentzian metric on spacetime.  We will conclude by arguing that there are significant costs to this response to the paradox.

We begin with some technical preliminaries.  In what follows, a \emph{(relativistic) spacetime} is a pair $(M,g_{ab})$, where $M$ is a smooth, Hausdorff, paracompact four-dimensional manifold and $g_{ab}$ is a smooth, non-degenerate metric of Lorentz signature $(+,-,-,-)$.  We say that a vector $\xi^a$ at a point $p$ of $M$ is \emph{timelike} if $g_{ab}\xi^a\xi^b > 0$; \emph{spacelike} if $g_{ab}\xi^a\xi^b<0$; and \emph{null} if $g_{ab}\xi^a\xi^b=0$.  A curve $\gamma:I\rightarrow M$ is \emph{timelike} if its tangent vector $\xi^a$ is at each point of its image; likewise for spacelike or null curves.  A vector (resp. curve) is \emph{causal} if it is timelike or null.  Finally, we say that a spacetime $(M,g_{ab})$ is \emph{temporally orientable} if there exists a continuous timelike vector field on $M$; a \emph{temporal orientation} is a choice $\eta^a$ of some such vector field.  Given a temporal orientation $\eta^a$, a causal vector $\xi^a$ at a point $p$ is \emph{future-directed} if $g_{ab}\eta^a\xi^b > 0$; causal curves are future-directed if their tangent vectors at each point are.  In what follows, we consider only temporally orientable spacetimes and suppose that we have fixed some temporal orientation.

To state the first assertion of the paradox, we need to introduce some notions related to the global (causal) structure of a spacetime \citep{Geroch+Horowitz, Hawking+Ellis, ManchakGS}.  Fix a spacetime $(M,g_{ab})$.  We define two two-place relations on the points of $M$.  First, for any $p,q\in M$, we have $p\ll q$ if there exists a future-directed timelike curve from $p$ to $q$; and second, we have $p< q$ if there exists a future-directed causal curve from $p$ to $q$.  The \emph{timelike future} of a point $p\in M$ is the set $I^+(p)=\{q:p\ll q\}$, and the \emph{causal future} of $p$ is the set $J^+(p)=\{q:p < q\}$.  Similar definitions yield the \emph{timelike} and \emph{causal pasts} of $p$, denoted $I^-(p)$ and $J^-(p)$, respectively.  The timelike future of a set $S\subseteq M$, meanwhile, is given by $I^+(S) = \cup_{p\in S}I^+(p)$; likewise for the sets $J^+(S)$, $I^-(S)$, and $J^-(S)$.

A point $p\in M$ is a \emph{future endpoint} for a future-directed causal curve $\gamma:I\rightarrow M$ if, for every neighborhood $O$ of $p$, there exists an $s'\in I$ such that for all $s\geq s'$, $\gamma(s)\in O$; likewise, \emph{mutatis mutandis}, for \emph{past endpoint}. The \emph{future domain of dependence} of $S$, $D^+(S)$, is the set of points $p\in M$ such that every causal curve with future endpoint $p$ and no past endpoint intersects $S$; likewise for the \emph{past domain of dependence}, $D^-(S)$.  The \emph{domain of dependence}, $D(S)$, is the set $D^+(S)\cup D^-(S)$. A subset $S\subset M$ is \emph{achronal} if $I^+(S)\cap S=\emptyset$.  Finally, a \emph{Cauchy surface} $\Sigma$ is a closed, achronal set with the property that $D(\Sigma)=M$.

Cauchy surfaces have the feature that the entire manifold lies in their domain of dependence, which means that any causal curve without endpoint passing through any event in the spacetime will necessarily intersect $\Sigma$.  For appropriate hyperbolic systems of differential equations, including Einstein's equation and Maxwell's equations, Cauchy surfaces have the property that initial data defined thereon determine unique global solutions.  For this reason, they are associated with a notion of \emph{determinism} in a spacetime, whereby the world at all times is ``determined'' by its state at one ``time'', i.e., on some Cauchy surface \citep{EarmanPD}.  A spacetime admitting a Cauchy surface $\Sigma$ is called \emph{globally hyperbolic}; if $(M,g_{ab})$ is globally hyperbolic, then it admits an entire foliation by Cauchy surfaces, in the sense that $M$ is diffeomorphic to the product manifold $\Sigma\times \mathbb{R}$, where each copy of $\Sigma$ turns out to be a Cauchy surface \citep{GerochSplitting,Bernal+Sanchez}.  Thus, we may think of a globally hyperbolic spacetime as consisting of a sequence of ``complete'' states, parameterized by global time.  We can then think of ``Cauchy evolution'' as consisting of transitions between such states over time.

We can now state the first assertion of the paradox, which is a version of the so-called ``cosmic censor[ship] hypothesis (CCH)'' \citep[p. 204]{Wald}.\footnote{See also \citet{Penrose}.}
\begin{quote}
\begin{itemize}
\item[(CCH)]\label{CCH} All physically reasonable relativistic spacetimes are globally hyperbolic.
\end{itemize}
\end{quote}
By ``physically reasonable'', here, we mean that the spacetime represents the global history of some possible physical process.  The idea is that general relativity is a permissive theory, in the sense that there are many models that are allowed by the theory, but which we might rule out of consideration on other grounds, such as our background understanding of other physical laws.

This condition is very strong: it is easy to come up with examples of spacetimes that are generally well-behaved, but which are not globally hyperbolic \citep[Ch. 2]{Earman}.  But for many physicists, (CCH) is axiomatic.  For instance, as we noted above, global hyperbolicity is sufficient, in the presence of background assumptions about the hyperbolic character of the laws governing matter evolution, for global determinism; if one is committed to the universe being deterministic, then (CCH) is a natural assertion to accept.\footnote{It is interesting to ask whether global hyperbolicity is also \emph{necessary} for global determinism.  The answer is apparently ``no,'' but further work is called for.  We are grateful to David Malament for raising this question.}  It rules out the possibility of ``naked'' singularities, i.e., singularities from which causal processes may emanate and influence distant events, the possibility of which many physicists take to be pathological and unrealistic.\footnote{Naked singularities are to be contrasted with singularities such as those in a black hole, which are ``hidden'' by an event horizon and therefore cannot influence distant events.}  And perhaps most importantly, as Maudlin himself emphasizes, global hyperbolicity appears to be necessary for global unitary evolution in the context of quantum theory set in curved spacetimes \citep{Belot+etal}.  Insofar as many physicists take unitary evolution to be the hallmark of quantum theory, giving up on (CCH) is a significant cost.

\begin{figure}[h]    \centering
   \includegraphics[width=3in]{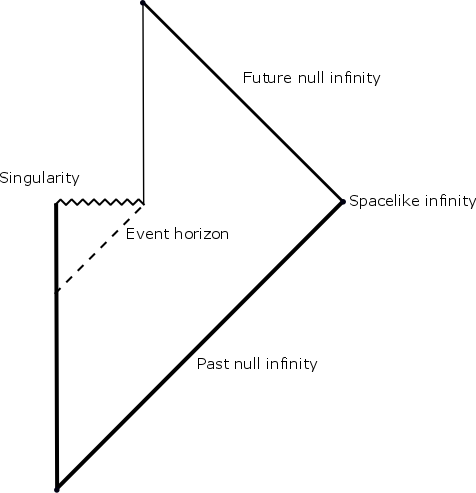}
   \caption{\label{PenroseBHE1} The Penrose diagram representing an evaporating black hole.  There is a (spacelike) singularity behind an event horizon, as in the Schwarzschild solution (say), but there also exist events ``after'' the black hole has evaporated, i.e., events such that no future-directed causal curve passing through those events encounters the singularity (or the event horizon).}
\end{figure}

The second assertion has a somewhat different character.  It concerns a particular class of spacetimes, which we will call \emph{evaporation spacetimes}, representing the physical process known as black hole evaporation \citep{HawkingR1,HawkingR2}.  These are spacetimes exhibiting (some of) the qualitative features of the spacetime whose Penrose diagram is depicted in Fig. \ref{PenroseBHE1}.  In such a diagram, the causal structure of spacetime is clear; null geodesics are represented by straight lines at $\pm 45^\circ$.\footnote{See \cite{Hawking+Ellis} for details concerning Penrose diagrams.} In an evaporation spacetime, one has a black hole which is a (spacelike) singularity that lies behind an event horizon.  But the black hole gradually evaporates and disappears, which means that it does not persist forever: its event horizon does not extend to null infinity.  This means that there exist events that occur ``after'' the black hole has ceased to exist, in the sense that there are points $p\in M$ such that the entire event horizon lies in the (closure of) $J^-(p)$. If one thinks of an observer following a timelike worldline far from the event horizon, there will be a period in that observer's history during which it would be possible, were she to change course, to eventually pass behind the event horizon and fall into the black hole; but at some (proper) time, that will no longer be possible, because no future-directed timelike curves intersecting her worldline at that time meet the singularity.

\begin{figure}[h]    \centering
   \includegraphics[width=6in]{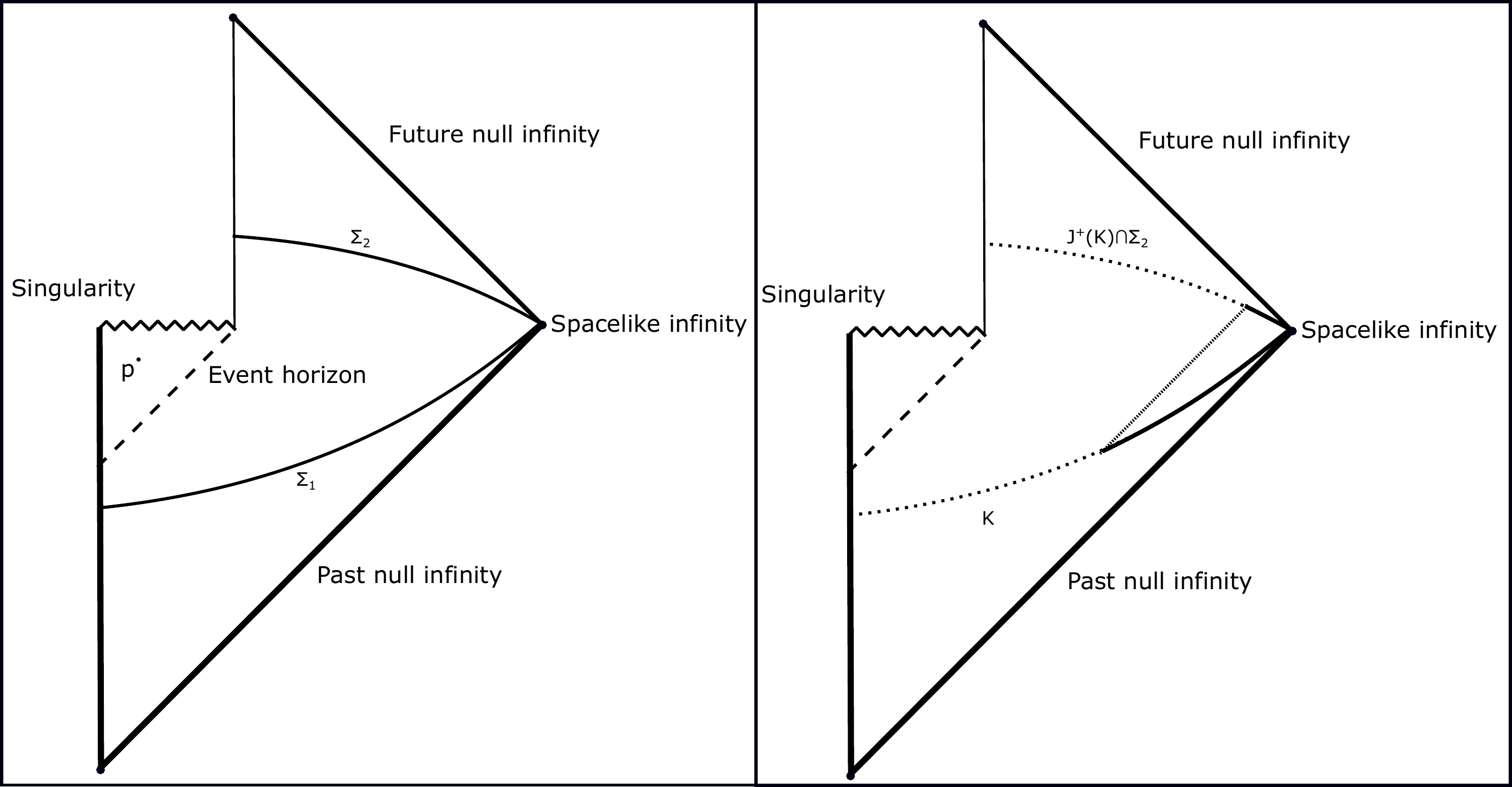}
   \caption{\label{PenroseBHE2} The same spacetime as represented in Fig. \ref{PenroseBHE1}.  In (a), closed achronal sets $\Sigma_1$ and $\Sigma_2$ are depicted.  Here $\Sigma_1$ is connected and $\Sigma_2$ is edgeless.  The point $p$ is such that $p\in D^+(\Sigma_1)$ and $p\not\in J^-(\Sigma_2)\cup J^+(\Sigma_2)$.  In (b), the sets $K=\Sigma_1-D^-(\Sigma_2)\cap \Sigma_1$ and $J^+(K)\cap\Sigma_2$ are labeled.}
\end{figure}

We make the notion of an evaporation spacetime precise as follows.  First, we need a little more background.  The \emph{edge} of a closed, achronal set $S\subset M$ is the set of points $p\in S$ such that every open neighborhood $O$ of $p$ contains points $q\in I^+(p)$, $r\in I^-(p)$, and a timelike curve from $r$ to $q$ that does not intersect $S$. A {\em slice} is a closed, achronal surface without edge. We say that a time-orientable spacetime $(M,g_{ab})$ is an \emph{evaporation spacetime} if there is a point $p \in M$ and slices $\Sigma_1, \Sigma_2 \subset M$ with $\Sigma_1$ connected such that: (a) $p\not\in J^-(\Sigma_2)\cup J^+(\Sigma_2)$; (b) $p\in D^+(\Sigma_1)$ and the set $K_1=\Sigma_1-D^-(\Sigma_2)\cap \Sigma_1$ is such that $J^+(K_1)\cap\Sigma_2$ has compact closure; and (c) for every other connected slice $\Sigma \subset M$ with $p\in D^+(\Sigma)$, the set $K=\Sigma-D^-(\Sigma_2)\cap \Sigma$ is such that $J^+(K)\cap\Sigma_2$ has compact closure.\footnote{To be explicit, we take this condition to be \emph{necessary} for being an evaporation spacetime, but do not take a stand on whether it is sufficient.}  As one can easily see from Fig. \ref{PenroseBHE2}, the spacetime we have been considering meets these conditions.  The surfaces $\Sigma_1$ and $\Sigma_2$ are chosen so that $\Sigma_1$ slices the spacetime ``below'' the event horizon; while $\Sigma_2$ slices it ``above'' the event horizon.  The point $p$ mentioned in the definition may then be any point at all lying behind the event horizon.  The region $K$ is the part of $\Sigma_1$ that is not (fully) determined by what occurs on $\Sigma_2$, in the sense of ``determined'' described above; the stated condition (b) on $K$ is a way of capturing the idea that the ``causal strangeness'' in the spacetime is confined to a bounded region.

Given this, the second assertion may be stated as follows.
\begin{quote}
\begin{itemize}
\item[(BHE)]\label{BHE} Some evaporation spacetimes are physically reasonable.
\end{itemize}
\end{quote}
Why accept (BHE)?  We will not rehearse the arguments for Hawking radiation or black hole evaporation here.  We remark only that there appear to be compelling theoretical grounds, arising from quantum field theory in curved spacetime, for believing that black holes radiate---and that it is a consequence of that radiation that they ultimately evaporate \citep{HawkingR1,HawkingR2,UnruhBHE,Birrell+Davies,Fulling,WaldQFTCST}.
This is not dispositive, of course: no one has observed Hawking radiation from a black hole---much less black hole evaporation.  It is possible that Hawking radiation does not occur, or that it is impossible for black holes to evaporate.  But in either case, it would seem to follow that there is a profound error in our understanding of quantum field theory and/or black holes.

We are now in a position to state the paradox.  It is captured by a theorem due to \citet{Kodama} (see also \citet{Wald1984}):
\begin{thm} No evaporation spacetime is globally hyperbolic.\end{thm}
It is a consequence of this theorem that (CCH) and (BHE) are incompatible: one cannot accept both the Cosmic Censorship hypothesis, at least in this form, and also believe that evaporation spacetimes are physically reasonable.  This is what we take to be the paradox.  To resolve it, something must be given up.

\begin{figure}[h]    \centering
   \includegraphics[width=3in]{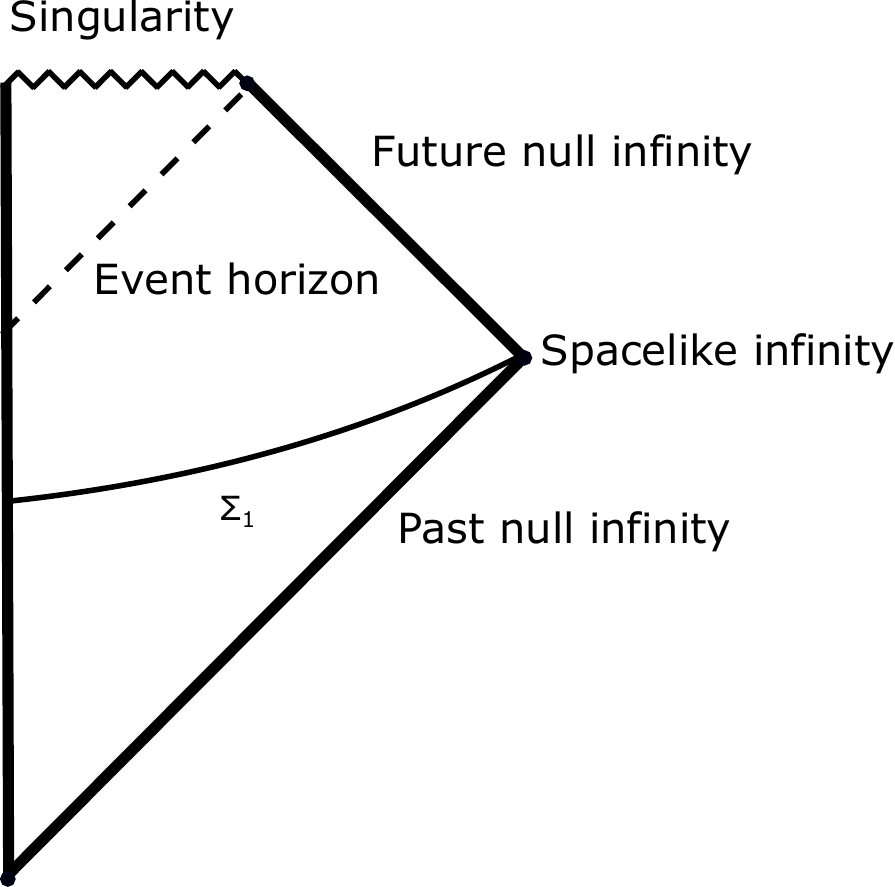}
   \caption{\label{BottomHalf} The bottom half of the spacetime depicted in Fig. \ref{PenroseBHE1}. Here $\Sigma_1$ is a Cauchy surface and the spacetime is globally hyperbolic.  But this spacetime can be extended, and its extension to that of Fig. \ref{PenroseBHE1} is not globally hyperbolic.  (Note that this spacetime is conformally equivalent to a Schwarzschild spacetime, but depicts a different situation.)}
\end{figure}

To get a handle on why this situation is troubling, consider the following.  Focus attention on just the bottom half of Fig. \ref{PenroseBHE1}, as depicted in Fig. \ref{BottomHalf}.  In this spacetime, the achronal set $\Sigma_1$ \emph{is} a Cauchy surface and the spacetime is globally hyperbolic.  Thus, there is a sense in which what happens on $\Sigma_1$ determines what happens in the entire spacetime---at least if we suppose that the Hawking radiation emitted by the black hole satisfies a suitable system of differential equations.\footnote{In what follows, we are, by necessity, somewhat impressionistic, because we do not know how to characterize the laws governing Hawking radiation classically, in such a way that the evaporation spacetime, including the Hawking radiation, is a solution to some system of equations.  But we hope this discussion captures the flavor of what is going on, even if it cannot be made perfectly precise.}  One can think of $\Sigma_1$ as a complete specification of the universe at a time, representing a black hole in the process of emitting radiation; the spacetime in Fig. \ref{BottomHalf} represents a sequence of such states over time, evolving in a lawlike manner, corresponding to  (the initial proper segment of) a physical process that we have good reason to believe occurs in our own universe.  Indeed, basic existence and uniqueness results for hyperbolic systems, including various Einstein-matter systems, suggest that we should expect there to be a unique maximal Cauchy development of $\Sigma_1$ \citep{GerochPDE,Rendall}: that is, there should be a maximal spacetime that we get by allowing $\Sigma_1$ to evolve according to the laws---and indeed, its Penrose diagram should look like Fig. \ref{BottomHalf}.   But the spacetime in Fig. \ref{BottomHalf} is, by construction, extendible, in the sense that there is a proper isometric embedding of this spacetime into the one depicted in Fig. \ref{PenroseBHE1}.  Thus, there is a sense in which the process can continue beyond its maximal Cauchy evolution.  And insofar as the physically correct extension of the spacetime in Fig. \ref{BottomHalf} is the spacetime in Fig. \ref{PenroseBHE1}, it follows that the laws, plus initial data specified on $\Sigma_1$, do \emph{not} determine what happens indefinitely into the future, because this second spacetime is not globally hyperbolic, and $\Sigma_1$ fails to be a Cauchy surface.  One reaches a horizon across which the laws no longer determine what happens.  This situation suggests that something has gone seriously wrong.  Hence the paradox.

There are many responses to this paradox available in the literature.  Most responses involve rejecting (BHE): one might argue, for instance, that black holes do not actually form \citep{Fuzzball}, or that they form and radiate, but never completely evaporate \citep{EllisRemnant}.  In such cases, one effectively denies that evaporation spacetimes are physically reasonable, by arguing that some other physical process becomes operative at some relevant scale.  Much less common, but defended by \citet{Unruh+Wald}, is to reject (CCH): that is, to accept that evaporation spacetimes are physically reasonable, and likely represent physical processes in our own universe, and accept that this leads to some surprising physical consequences, such as the failure of unitarity in evolution from $\Sigma_1$ to $\Sigma_2$; and the failure of retrodictability from $\Sigma_2$.

Maudlin, however, does not reject either of the premises: he apparently maintains that both (CCH) and (BHE) are true.  How can this be, given the Kodama-Wald theorem above?  As Maudlin presents it, the ``solution'' to the apparent paradox is to recognize (correctly) that $\Sigma_2$ in Fig. \ref{PenroseBHE2}(a) is not a Cauchy surface.  But this observation does not lead him to reject (CCH); instead, he argues that a \emph{different} surface, consisting in the union of $\Sigma_2$---now renamed ``$\Sigma_{2out}$''---and another achronal set, which he calls $\Sigma_{2in}$, together form a Cauchy surface.  The idea is that there is some critical surface, $\Sigma_{crit}$, after which Cauchy evolution proceeds along \emph{disconnected} Cauchy surfaces.  There is no failure of determinism, or of unitarity; rather, one just needs to take account of the part of the global state corresponding to $\Sigma_2$ that remains trapped behind the event horizon.

\begin{figure}[h]    \centering
   \includegraphics[width=3in]{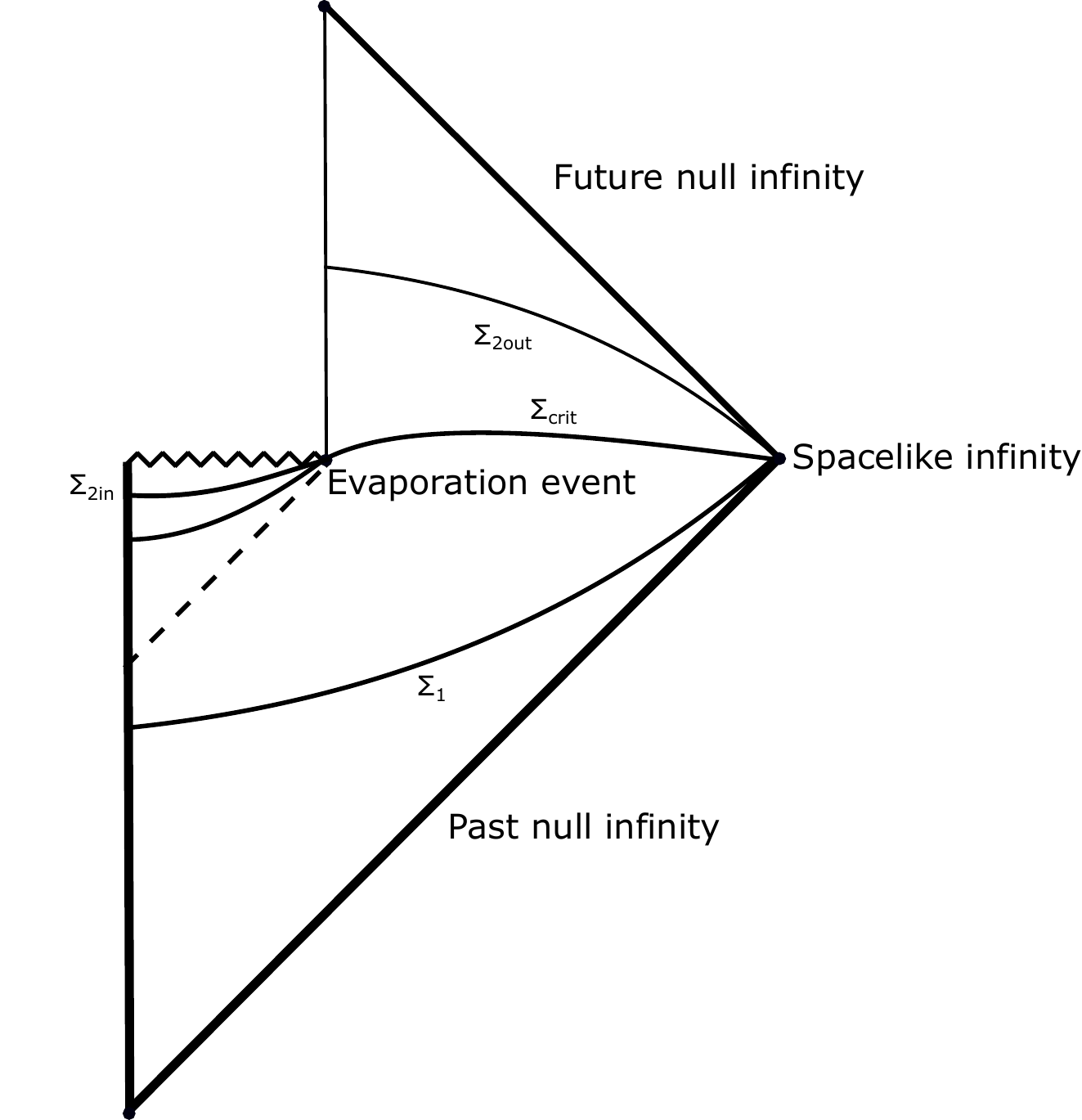}
   \caption{\label{MaudlinFig}}
\end{figure}

Of course, the ``solution'' just sketched cannot be the whole story, because it is flatly inconsistent with the Kodama-Wald theorem.  After all, that theorem says that evaporation spacetimes cannot be globally hyperbolic; thus, $\Sigma_{2in}\cup\Sigma_{2out}$ is no more a Cauchy surface than $\Sigma_2$ was (or $\Sigma_1$ is).  Maudlin does not address this point---but he does address an inconsistency between his solution and another theorem, and his remarks there suggest what is really going on.  He writes (in connection with the Geroch splitting theorem \citep{GerochSplitting}),
\begin{quote}
The	solution to	this puzzle	can	easily escape one's notice.	Geroch's proof,	like most proofs in	this field,	begins by characterizing a	space-time as a “4-dimensional manifold	with a	smooth metric $g_{ab}$ of signature	$(+,-,-,-)$.  But our evaporating black	hole space-time	is not a manifold. (p. 12)
\end{quote}
In other words, Maudlin is drawing attention to a third, suppressed premise in the paradox as described above, which is assumed in both Kodama and Wald's proofs of the Kodama-Wald theorem.  It may be stated as follows:
\begin{quote}
\begin{itemize}
\item[(RST)] Any physically reasonable relativistic spacetime may be represented by a smooth (Hausdorff, paracompact) manifold without boundary, with a Lorentz signature metric.
\end{itemize}
\end{quote}
It is (RST) that Maudlin rejects.

We do not deny that there may be good reasons to reject (RST).  After all, the arguments in favor of (BHE) rely on a semi-classical analysis according to which spacetime is treated classically and the radiation is treated quantum field theoretically.  One might well expect that a full understanding of Hawking radiation will wait on a theory of quantum gravity---and on several approaches to quantum gravity, the description of space and time as a smooth manifold breaks down, because space and time turn out to be fundamentally discrete.  (Maudlin himself does not offer an argument along these lines.)  But whatever else is the case, observing that (RST) may be rejected does not simply dissolve the paradox.\footnote{Or as Maudlin puts it: \begin{quote} ...there is no problem that needs to be solved. Once one is careful and precise about the	principles at issue, such as retrodictability and unitarity and the meaning of	information	conservation, one finds	that there is no reason	to think they were ever	violated in	the	first place. (p. 24)\end{quote}}  As we will now argue, there are serious costs associated with giving this premise up (and giving it up in the particular way that Maudlin does) that bear reflection.  Indeed, there are good reasons to think that Maudlin's rejection of (RST) yields at best a pyrrhic victory.

One might reject (RST) in a variety of ways. For example, one might replace (RST) with a weaker but analogous premise (RST)* that just relaxes some combination of the conditions on the underlying manifold.  The Hausdorff condition in particular is often dropped and relativistic spacetime structure can be investigated in this generalized framework \citep{Hajicek1971,Hawking+Ellis, Earman2008}. One might also want to consider manifolds with boundary \citep{Geroch+Horowitz}. There are drawbacks to representing spacetime using non-standard manifolds such as these, but they are clearly defined and legitimate objects of study. One can, for example, keep track of which of the standard theorems of general relativity carry over to any one of the non-standard arenas and which do not. One could even construct arguments in favor of non-standard spacetimes on physical grounds. Fine. At this stage, we simply wish to emphasize that this does not seem to be what Maudlin is up to; the manifold-like structure he considers is neither a (Hausdorff or non-Hausdorff) manifold with boundary nor a manifold without boundary.\footnote{Maudlin himself states that his structure is not a manifold without boundary (2017, p. 12). And we know that any $n$-dimensional manifold with boundary has a boundary of dimension $n-1$ \citep[p. 12]{Hawking+Ellis}. Since the evaporation event is ``a single point'' \citep[p. 20]{Maudlin}, it cannot be the boundary of a four-dimensional spacetime.} If Maudlin's manifold-like structure is not a manifold without boundary and not a manifold with boundary, what kind of structure is it? What are its properties? Maudlin never tells us. Indeed, we know almost nothing about the proposed manifold-like structure and the spacetime constructed from it except that it amounts to ``adding'' a point to a standard spacetime structure.

One might wonder \citep[cf.][p. 20]{Maudlin}: How much difference can adding  a single point make? It turns out quite a bit; sometimes adding a point to a standard spacetime structure can be logically incoherent if we require (as Maudlin seems to want to do) that a smooth Lorentzian metric is everywhere defined.

\begin{figure}[h]    \centering
   \includegraphics[width=6in]{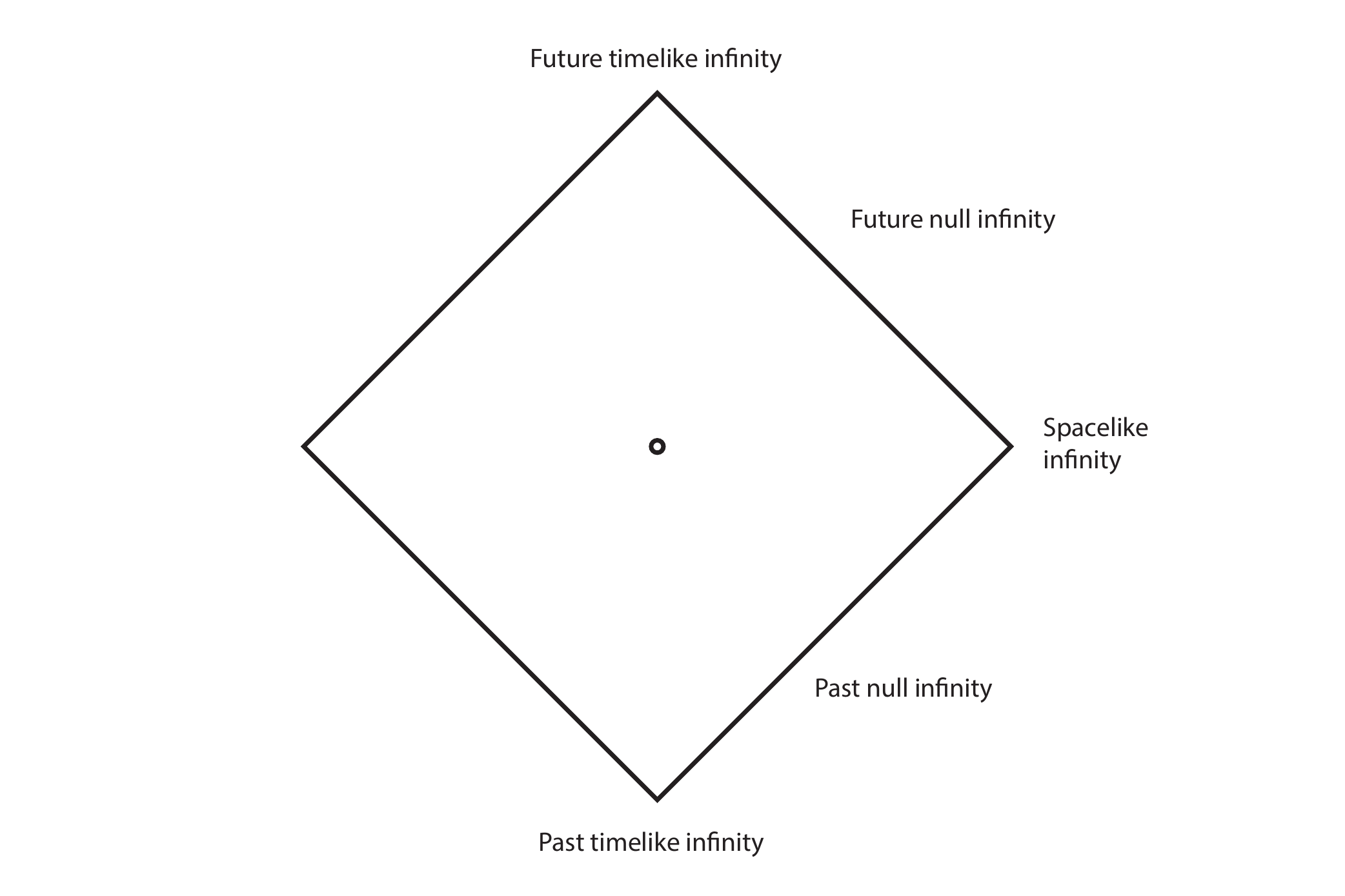}
   \caption{\label{ConMink} The Penrose diagram for both $(M,\eta_{ab})$ and $(M,\Omega^2\eta_{ab})$. One can add back the ``missing point'' in the former spacetime but not the latter.}
\end{figure}

First, consider two-dimensional Minkowski spacetime $(\mathbb{R}^2, \eta_{ab})$ in standard $(t,x)$ coordinates where $\eta_{ab}=\nabla_at\nabla_bt-\nabla_ax\nabla_bx$. Now remove the point $(0,0)$ and let the resulting manifold be $M$. Next, consider the spacetime $(M,\eta_{ab})$. Of course we could put the ``missing point'' back in $(M,\eta_{ab})$ and the resulting structure would be a perfectly fine model -- Minkowski spacetime. But now consider the spacetime $(M,\Omega^2\eta_{ab})$ where $\Omega: M \rightarrow \mathbb{R}$ is the function $\Omega(t,x)=1/(t^2+x^2)$. Clearly, the metric $\Omega^2\eta_{ab}$ blows up as one approaches the ``missing point''; one cannot put this point back in because there is no way to smoothly (or even continuously) extend the metric to it. So it seems that sometimes it makes perfect sense to add a point to a standard spacetime structure and sometimes it doesn't. What we take to be striking is that because $(M,\eta_{ab})$ and $(M,\Omega^2\eta_{ab})$ are conformally related, they have exactly the same Penrose diagrams (see Fig. \ref{ConMink}). Thus, it is not possible to determine if it is permitted to add a point to a standard spacetime structure simply by looking at the Penrose diagram of that structure. As we will argue presently, this point is important to understanding the pathology in the case of an evaporation spacetime.

\begin{figure}[h]    \centering
   \includegraphics[width=5in]{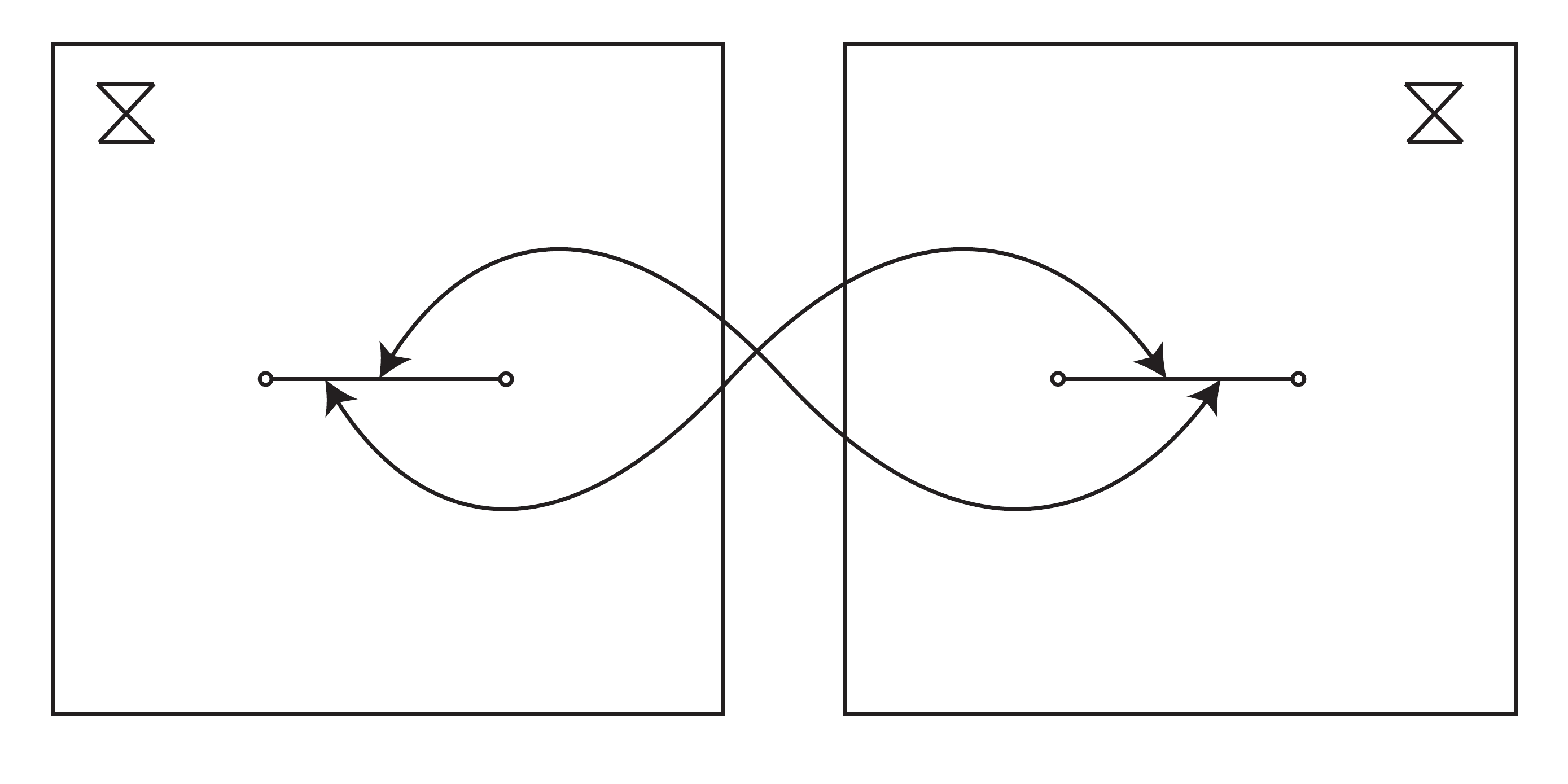}
   \caption{\label{Slits} The spacetime $(M,g_{ab})$. The ``top side'' of one slit is identified with the ``bottom side'' of the other and vice versa. One cannot add back any of the four ``missing points'' in a coherent way.}
\end{figure}

A second example illustrates the idea that even a ``well-behaved'' metric does not ensure that a point may be coherently added to a standard spacetime structure. Consider two copies of two-dimensional Minkowski spacetime $(\mathbb{R}^2, \eta_{ab})$ as defined above. Next cut a slit $\{(t,x): t=0, |x|\leq 1\}$ out of the $x$-axis in each of the underlying manifolds. Finally, excluding the points $(0,-1)$ and $(0,1)$ in each copy, identify the ``bottom side'' of one slit with the ``top side'' of the other slit and vice versa (see Fig. \ref{Slits}). The resulting structure---call it $(M,g_{ab})$---is a standard spacetime. One might wonder if any one of the four excluded points can be added back into $(M,g_{ab})$. No: ``We cannot put these points back in because we were perverse enough to extend the top and bottom sides of the $x$-axis on different sheets'' \citep[p. 59]{Hawking+Ellis}. Essentially, what is going on is that $(M,g_{ab})$ is ``on the verge'' of pathological behavior and only the ``missing points'' are keeping such behavior at bay.

What these examples show is that although it may \emph{seem} as if it is a trivial matter to extract or add a single point to a spacetime---and that the physics cannot possibly depend on whether a single point is present or not---in general this is simply not true.  In fact, a ``missing point'' often signals more general pathological behavior in some neighborhood of the manifold.  And indeed, that is precisely what is going on in the case of evaporation spacetimes.

\begin{figure}[h]    \centering
   \includegraphics[width=3in]{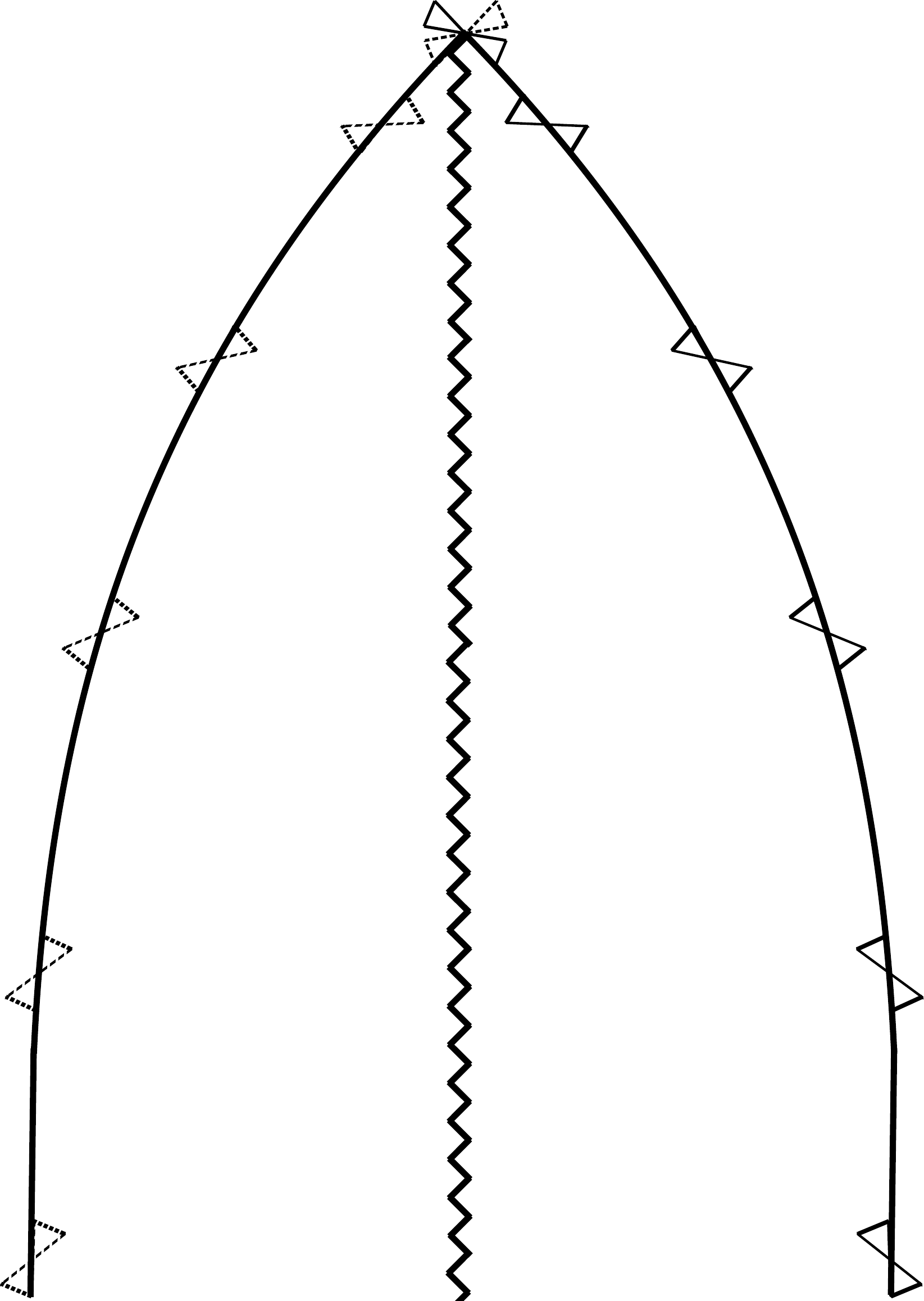}
   \caption{\label{WaldFig} A spacetime diagram depicting an evaporating black hole; compare with \citet[Fig. 1]{Wald1984}.  Here we see that the causal cones along the event horizon tend to rotate in opposite directions as one moves up the diagram, leading to an inconsistency at the top, i.e., the ``evaporation event''.  Note that if we assume---as physicists often do---that the spacetime ``after evaporation'' is approximately Minkowski spacetime, we see that the in consistency becomes even worse.}
\end{figure}

To see this point most clearly, it is best to look at a different representation of a collapsing black hole, because as we saw in the first of the examples discussed above, Penrose diagrams can be misleading with regard to whether a point may be ``added'' to a spacetime.  In figure \ref{WaldFig}, we show a standard spacetime diagram for an evaporating black hole.  (This diagram is very similar to \citet[Fig. 1]{Wald1984}.)  As \citet[p. 162]{Wald1984} says, this diagram ``accurately shows that after the black hole forms from the collapse of matter, it shrinks in size as it radiates until it disappears altogether.''  It also depicts the tendency of ``small'' black holes to evaporate more rapidly than ``large'' ones, because the event horizon curves towards the singularity as one approaches the top of the diagram.

In this diagram, we see that as one approaches the ``evaporation event'' (i.e., the point at the top of the cone) from different directions, the light cones tend to tilt in incompatible ways.  There results a failure of continuity of the light cone structure at the top of the diagram, i.e., at the evaporation event.  The upshot is that no causal cone can be consistently defined at the evaporation event.  Ultimately, that there is no way to define a consistent causal structure here is what the Kodama-Wald theorem captures.

If we compare Fig. \ref{WaldFig} with Fig. \ref{MaudlinFig}, we see that the causal pathologies at the ``evaporation event'' are obscured in the Penrose diagram.  This is because, at each point, a conformal transformation is performed to make the null directions point at 45 degree angles from the vertical.  But as we saw in the first example above, of conformally transformed Minkowski spacetime, the Penrose diagram does \emph{not} depict any information about the character or magnitude of the conformal transformation at different points, which in turn means that it does not provide sufficient information to determine whether a point can be consistently added to the original spacetime.  In the present case, the event horizon is, in fact, ``shrinking'' over time.  But this fact is manifestly \emph{not} represented in the Penrose diagram, precisely because Penrose diagrams do not accurately represent distance relations.  In fact, the Penrose diagram depicts the event horizon as ``expanding'' (at the speed of light).  Hence we see that in order to present the spacetime as having a well-behaved conformal structure, one needs to apply a conformal transformation that makes a parameter (roughly the ``size'' of the event horizon) appear to grow when in fact it approaches $0$.  This is strikingly similar to the Minkowski spacetime example already given, and shows why the Penrose diagram gives a misleading picture of the causal structure at the evaporation event.

The breakdown of causal structure at the ``evaporation event'' has consequences.  For one, it leads to a breakdown of the laws of physics.  Consider, for instance, Maxwell's equations.  These are a hyperbolic system whose associated ``causal cone'' always agrees with the metric lightcone at each point \citep{GerochPDE}.  If no consistent metric lightcone can be defined at a point, it follows that Maxwell's equations, as standardly understood, also cannot be defined there.  More generally, the metric is essential for expressing known laws of physics; if the metric cannot be defined at a point, it is not clear that any laws can be defined there.  Thus, even if the spacetime-like structure Maudlin proposes \emph{is} in some sense globally hyperbolic, it is hard to see how any notion of global determinism (or retrodictability) obtains, given that the laws of nature apparently breakdown somewhere in spacetime.

One might take the arguments just given to weigh against merely adding a single point-like ``evaporation event'' to an evaporation spacetime, at least if one wishes to have a well-defined causal structure.  But perhaps there is a different solution available: one might consider either adding multiple points to the spacetime, or else adding a point and then attempting to modify the spacetime in a neighborhood of that point.  This second option would be analogous to an operation one can perform on a cone.  A cone is usually defined as a flat, two dimensional Riemannian manifold, without vertex.  One may think of this as a manifold with a single point ``removed''.  One cannot simply ``add'' the vertex back in, however, as the curvature at that point would not be defined (it would, effectively, be infinite).  One can, however, ``add'' the point by smoothing out the top of the cone.  To do so, one introduces non-vanishing curvature in some neighborhood of the vertex of the cone, so that one does not ``merely'' add a point.  One might expect that some similar sort of manipulation could allow one to both ``add'' an evaporation event and also avoid the causal misbehavior we have already discussed---perhaps without giving up on (RST).  But it is here that we see the power of the Kodama-Wald theorem, once again: whatever else is the case, no matter how one added the point and modified the spacetime in a neighborhood of that point, the resulting spacetime could only be globally hyperbolic if it were no longer an evaporation spacetime.

Given the arguments above, it would seem that one cannot merely add an ``evaporation event'' to an evaporation spacetime to resolve the paradox in a satisfactory way, at least if one demands a suitable Lorentz-signature metric to be definable everywhere in spacetime. But let us read Maudlin as charitably as we can and suppose there is some way to make things work out. There remains a serious problem.\footnote{Our thanks to Bob Wald (personal correspondence) for the following argument.  Wald also points out a second feature of the spacetime Maudlin considers, which is that it is not clear how to continue past-directed null geodesics approaching the evaporation event from above: both continuing along the event horizon and continuing directly to past null infinity would be compatible with the causal structure implied by Maudlin's Penrose diagram (with the evaporation event).  This argument is not as strong as the one in the main text, but it helps to further emphasize the causal pathology at the evaporation event.} Consider a past-directed timelike geodesic at $r=0$ (i.e., the axis of symmetry for the diagram) with past endpoint at the evaporation point. Note that given the spherical symmetry of the evaporation spacetime, the geodesic must be invariant under rotations. Now, either the geodesic can be extended (as a geodesic) through the evaporation point into the past or it cannot be. If we suppose the former, then the geodesic cannot remain at $r=0$ because of the singularity. But if the the geodesic enters the $r> 0$ region, it must break the spherical symmetry of the of the spacetime: impossible. Now suppose the geodesic cannot be extended through the evaporation event. In this case, the past-inextendible timelike geodesic terminates at the evaporation event and never reaches Maudlin's surface $\Sigma_1$. It follows that $\Sigma_1$ is not a Cauchy surface and the spacetime fails to be globally hyperbolic.

So it seems that (1) there cannot be a Lorentizian metric defined at the evaporation point in Maudlin's nonstandard model and (2) even if there were, the model would fail to be globally hyperbolic. We close with one final point: (3) even if Maudlin's model were considered ``globally hyperbolic'' in some sense, it would fail to have some of the key properties of globally hyperbolic spacetimes.  The idea is that the usual physical and mathematical significance of global hyperbolicity, including that globally hyperbolic spacetimes are always causally well behaved and support global existence and uniqueness results for hyperbolic systems, follows from where it sits in a whole networks of properties a spacetime can have.  In particular, in the case of ordinary spacetime, global hyperbolicity sits at the top of the causal hierarchy.  But as we will presently show, that is no longer true, if what we mean by ``global hyperbolicity'' has been extended to include something like the spacetime Maudlin considers.  It follows that even if one can make sense of a notion of ``global hyperbolicity'' that includes the spacetime Maudlin proposes (with evaporation event), it is not clear what significance that has.  In a sense, this should not be a surprise: Maudlin himself acknowledges that standard theorems fail for the non-manifold structures he considers.  But we think it is valuable to reflect on just what is being given up.

In standard presentations of the causal structure of spacetime, one sometimes encounters an early lemma \citep[p. 183]{Hawking+Ellis}.\\

\begin{lem} For all spacetimes $(M,g_{ab})$, and for all points $p, q, r$ in $M$, if $q$ is in $J^+(p)$ and $r$ is in $I^+(q)$, then $r$ is in $I^+(p)$. \end{lem}

One direct consequence of the lemma is the following: For all spacetimes $(M,g_{ab})$, and for all points $p \in M$, the closure of $I^+(p)$ is identical to the closure of $J^+(p)$. This shows one sense in which the timelike and causal future of a point (in any spacetime) are ``almost'' the same. Now it turns out that $I^+(p)$ is always an open set. One might wonder: is $J^+(p)$ always closed? No: spacetime ``holes'' can sometimes prevent this. Indeed, Minkowski spacetime with one point removed from the manifold is standardly used to illustrate the idea \citep[p. 191]{Wald}. However, when one limits attention to globally hyperbolic spacetimes, one does find that $J^+(p)$ is always closed \citep[p. 208]{Wald}. Combining results, we have the following. \\

\begin{prop}For all globally hyperbolic spacetimes $(M,g_{ab})$, and for all points $p \in M$, the closure of $I^+(p)$ is $J^+(p)$. \end{prop}

\begin{figure}[h]    \centering
   \includegraphics[width=7in]{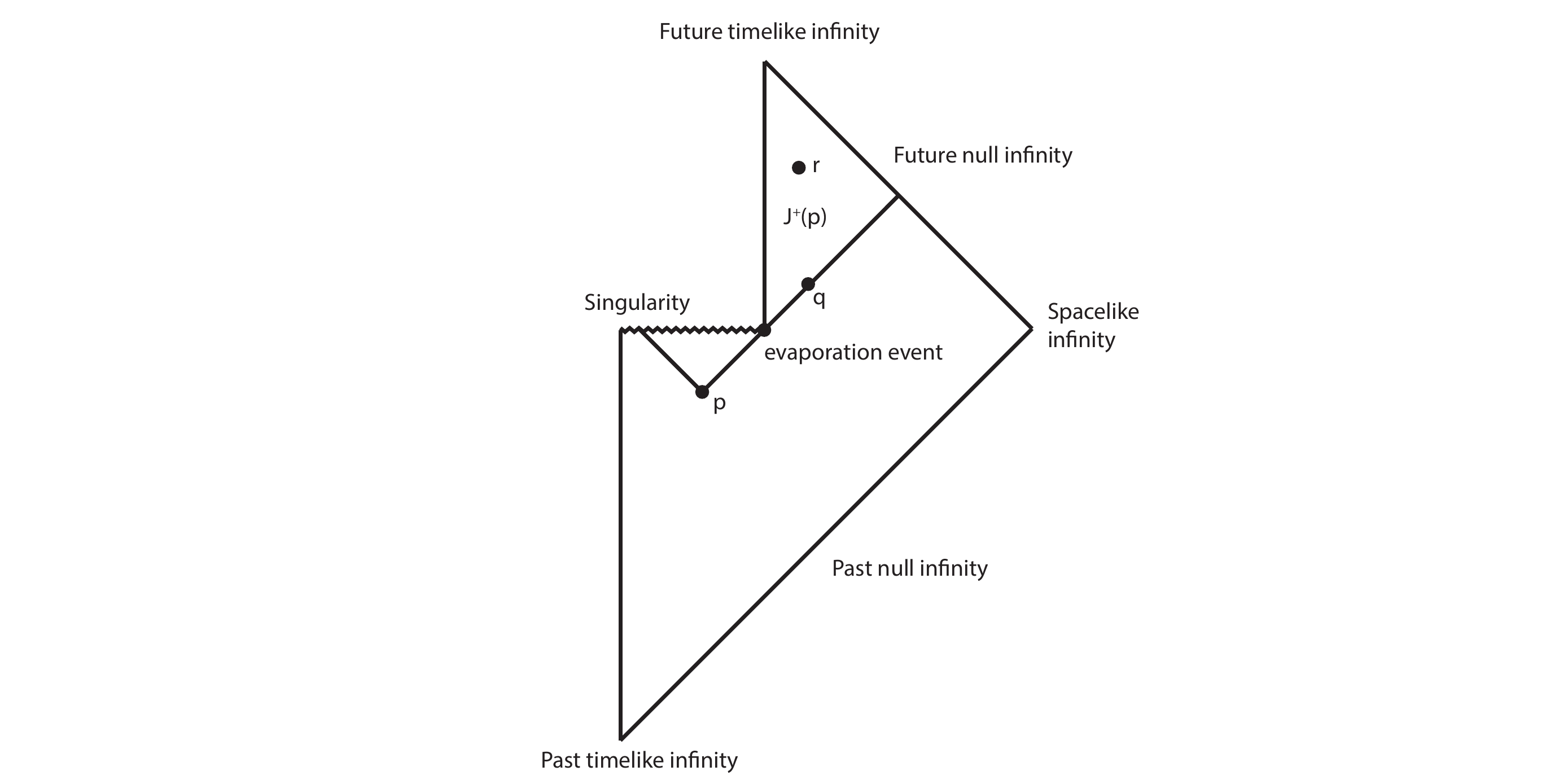}
   \caption{\label{CausCon} The Penrose diagram of Maudlin's model. We see that the model has a causal structure quite different from any standard spacetime since $q \in J^+(p)$ and $r \in I^+(q)$ but $r \notin I^+(p)$. Moreover, the model fails to be causally continuous since $I^-(p) \subseteq I^-(q)$ but $I^+(q) \nsubseteq I^+(p)$.  }
\end{figure}

Essentially, the proposition is one way of expressesing the idea that standard globally hyperbolic spacetimes are so causally well-behaved that no ``holes'' can be there. Now let us return to Maudlin's model with the ``evaporation event'' in place. Referring to the Penrose diagram and choosing any point $p$ on the event horizon, one finds that the above proposition comes out as false; the timelike future of the point is confined to the region inside the event horizon while the causal future of the point extends out past the ``evaporation event'' and reaches future null infinity. Given that the proposition fails on Maudlin's model, it would seems that the model's causal structure differs significantly from the causal structure of the standard globally hyperbolic spacetimes;  ``holes'' are permitted in this new context after all. What is more, it seems that these differences are rooted in the fact that that Maudlin's model has a causal structure unlike {\em any} standard relativistic spacetime. To see this, note that even the lemma fails on Maudlin's model (see Fig. \ref{CausCon}).

Not only is the causal structure of Maudlin's model quite unusual; it seems to lack fundamental properties of globally hyperbolic spacetimes. We have already touched on the notion of spacetime ``holes'' in the preceeding. Hawking and Sachs (1974) call them ``gaps'' instead. Let us investigate the concept in a bit more detail.

Recall that in standard presentations of causal structure, a hierarchy of causal conditions is central. One of these conditions is of special interest to us here: causal continuity. Roughly, a causally continuous spacetime is ``stable'' with respect to its causal structure and also has ``no really big gaps'' \citep[p. 288]{Hawking+Sachs}. Now, it turns out there is a difference between ``gaps'' and ``really big gaps'' and the difference will be important later on. We know that the causal hierarchy from causal continuity to global hyperbolicity can be thought of as a ``gap'' hierarchy \citep[p. 295]{Hawking+Sachs}. Globally hyperbolic spacetimes can have no ``gaps'' at all. Causally continuous spacetimes can have ``gaps'' but no ``really big gaps''. Minkowski spacetime with a point removed from the manifold illustrates the idea: it has ``gaps'' (it is not globally hyperbolic) but no ``really big gaps'' (it is causally continuous).

Formally, a spacetime $(M, g_{ab})$ is {\em causally continuous} if it is distinguishing (i.e. for all $p, q \in M$, if either $I^+(p)=I^+(q)$ or $I^-(p)=I^-(q)$, then $p=q$) and reflecting (i.e. for all $p, q \in M$, $I^-(p) \subseteq I^-(q)$ iff $I^+(q) \subseteq I^+(p)$). As a basic result, we have the following. \\

\begin{prop} Any globally hyperbolic spacetime is causally continuous. \end{prop}

The proposition tells us (unsuprisingly) that any spacetime with ``really big gaps'' cannot be globally hyperbolic in the standard sense. We take it to be significant that the model presented by Maudlin as ``globally hyperbolic'' renders this basic proposition false; the model contains the ``really big gaps'' prohibited not only by standard global hyperbolicity but also by (the comparatively weak condition of) causal continuity. To see this, take $p$ and $q$ to be as indicated in Fig. \ref{CausCon} above. Clearly, $I^-(p) \subseteq I^-(q)$ but $I^+(q) \nsubseteq I^+(p)$. Thus, we have shown (3) from above: even if Maudlin's model were considered ``globally hyperbolic'' in some sense, it would fail to have some of the key properties of globally hyperbolic spacetimes.

\end{document}